\documentstyle[twoside,fleqn,espcrc2,psfig]{article}

\newcommand{\Mz}{M_{\rm z}}
\newcommand{\cU}{{\cal U}}
\newcommand{\cZ}{{\cal Z}}
\newcommand{\cL}{{\cal L}}
\newcommand{\cR}{{\cal R}}

\newcommand{\AmS}{{\protect\the\textfont2
   A\kern-.1667em\lower.5ex\hbox{M}\kern-.125emS}}
\newcommand{\gt}{{\tilde g}}
\newcommand{\gtp}{{\tilde g}'}
\newcommand{\sint}{\sin^2{\theta_W}}

\title{The Weak Mixing Angle from an $SU(3)$ Symmetry at a TeV}

\author{Savas Dimopoulos\address{Physics Department, Stanford
University, Stanford, CA 94305-4060, USA.} and David
Elazzar Kaplan\address{SLAC, Stanford, CA 94025}
          }

\begin{document}

\begin{abstract}

  The measured values of
two electroweak gauge couplings appear to obey an
approximate (5\%) $SU(3)$ relation. Unless this is
an accident caused by fortuitous Planck-scale physics, it
suggests the presence of an $SU(3)$ symmetry near the
electroweak scale. We propose this to be a local $SU(3)$
which spontaneously ``mixes'' with $SU(2)
\times U(1)$ near a TeV. Although all the particles of the
standard model are SU(3)-singlets, this symmetry relates
the electroweak gauge couplings and can successfully
predict the weak mixing angle with a precision of a few
percent.  Since this mechanism operates at a TeV, it does not
require an energy desert and consequently can be embedded in
theories of TeV-gravity.

\vspace{1pc}
\end{abstract}

\maketitle

\noindent{\bf Introduction:}
The most impressive quantitative success of any extension of the
standard model is the ``unification'' prediction of the weak
mixing angle in the supersymmetric standard model (SSM) \cite{dg}:
that is, a correlation between $\alpha_s(M_Z)$ and
$\sint$, predicted in 1981 and confirmed by experiment
at the two percent level \cite{Ghilencea:2001qq} 
10 years later. This has made the SSM
the leading contender for physics beyond the standard model and
supports the reality of a supersymmetric energy desert extending
from a TeV to $10^{16}$ GeV. 

The reason for the enormity of the
desert is the large disparity between the measured value of
$\sint$ (= .231) \cite{Groom:in} at the electroweak scale and its
theoretical value at the fundamental (or unification) scale (= 3/8) 
\cite{GUTs,Pati:1974yy,Dimopoulos:1981yj}.
This disparity necessitates a large hierarchy between the
electroweak  and unification scales, to allow $\sint$
to evolve from  the experimental to its theoretical value.

In this paper we propose theories in which the theoretical value of
$\sint$ is predicted to be near its experimental value at $\Mz$ and
as a result have a fundamental scale near the electroweak energy --
and no desert.  Therefore in these theories a TeV-scale cutoff,
and therefore quantum gravity at a TeV \cite{Arkani-Hamed:1998rs},
can coexist with a precise prediction of the weak mixing angle.

A key observation is that the electroweak gauge couplings 
and some of the matter content of the
SM exhibit an approximate $SU(3)$ symmetry suggesting a more
fundamental gauge sector {\it at the weak scale}.  
The Higgs doublet is contained in a triplet of $SU(3)$ and weak
hypercharge is identified with the eighth generator of the $SU(3)$
\cite{Weinberg:1971nd,Pisano:ee}.  
The theory predicts $\sint=0.25$, close to the
measured value of $0.231$, at the scale of $SU(3)$
breaking.  The quarks, however, do not fit in this framework as
hypercharges smaller than $1/2$ in magnitude are impossible to
accommodate.  

In this letter, we first present a simple extension to the SM which
reproduces the approximate $SU(3)$ symmetry by correctly
predicting $\sint$ to the few percent level.  The well-known mechanism
\cite{Yanagida:1994vq,Weiner:2001pv} 
which we coin ``spontaneous mixing'' is used to
allow the SM sector to remain intact.  
We discuss the theoretical uncertainties in our model and compare to
those of supersymmetric GUTs.  
Next we show how the mechanism can be easily embedded 
in theories at the weak scale which predict charge
quantization, and we give two examples.  
Then we present an example of a model in which the
Higgs is a pseudo-Goldstone boson and the prediction of
$\sint$ is maintained.  We conclude with a brief discussion
of experimental signatures and compare with other approaches.

\noindent{\bf The Minimal Module:} Consider the SM; add to it a new gauge
group $SU(3)$ and a scalar $\Sigma$ which is a triplet of the new
$SU(3)$, has the SM quantum number of the Higgs,
$(2,-1/2)$ under $SU(2)\times U(1)$, and is a
singlet of ordinary color.  Let $\Sigma$ get a vacuum
expectation value (VEV) of the form:
\begin{equation}
\langle\Sigma\rangle = \pmatrix{M&0 \cr 0&M \cr 0&0}\, ,
\label{vev}
\end{equation}
breaking $SU(3)\times SU(2) \times U(1) \rightarrow SU(2) \times
U(1)$. (For an explicit example which produces this VEV, see the Appendix.)
At the scale $M$ the gauge couplings are related
by:
\begin{equation}
{1\over g^2} = {1\over g_3^2} + {1\over \gt^2}
\label{geq}
\end{equation}
and
\begin{equation}
{1\over g'^2} = {3\over g_3^2}
        + {1\over {\tilde g}'^2}\, ,
\label{gpeq}
\end{equation}
where $(g,g')$ are the gauge couplings of the standard electroweak
theory, $(\gt,\gtp)$ are the new $SU(2)\times U(1)$ couplings and 
$g_3$ is the $SU(3)$ gauge coupling.  All couplings
are evaluated at the scale $M$.  Note, $g'$ and ${\tilde g}'$ are
normalized such that the standard model particles have $U(1)$ charges
equal to their canonical hypercharges ({\it e.g.}, $Y=1/2$ for the 
lepton doublets). In the limit of large $\gt,\gtp$, 
both low energy gauge couplings 
$g$ and $g'$ are determined by the single $SU(3)$ coupling $g_3$.
Therefore, although the SM sector has no $SU(3)$ symmetry, the low
energy gauge couplings $g$ and $g'$ at the scale $M$ are related by
an (approximate) $SU(3)$ relation which leads to the value $\sint
\simeq .25$.  Using the renormalization group equations for the SM,
one finds $M_o\equiv M|_{\gt,\gtp\rightarrow\infty}$ to be
\begin{equation}
M_o = \Mz e^{- \frac{8\pi^2}{b - b'/3}(g^{-2} - g'^{-2}/3)}\, .
\label{schlep}
\end{equation}
Inserting the beta-function coefficients for the SM $(b,b')=(19/6,-41/6)$
and the value of the gauge couplings at $\Mz$, one finds $M_o=3.75$ TeV.
Therefore, in the exact $SU(3)$ limit, this is the prediction for the scale
of $SU(3)\times SU(2)\times U(1)$ breaking to $SU(2)\times U(1)$.

\noindent{\bf Theoretical Uncertainty:}
In this section we quantify the main theoretical uncertainties
in the module, namely the unknown values 
of the new gauge coupling $\gt$ and $\gtp$ and the scale $M$.
We then compare the status of this mechanism with that of normal
supersymmetric grand unified theories.

The biggest source of uncertainty comes from the couplings
${\tilde g}$ and ${\tilde g}'$.  If these couplings are
smaller than $g_3$, then the prediction for $\sint$
is completely washed out and $g$ and $g'$ are dominated
by the uncorrelated tilded gauge couplings.  On the other hand, if 
${\tilde g},{\tilde g}'$ are larger
than $g_3$ then there is a significant region in parameter space 
in which the $SU(3)$ group dominates the prediction for $\sint$
and corrections to the $SU(3)$-symmetric relation are of order a few \%. 

\begin{figure}
  \centerline{
    \psfig{file=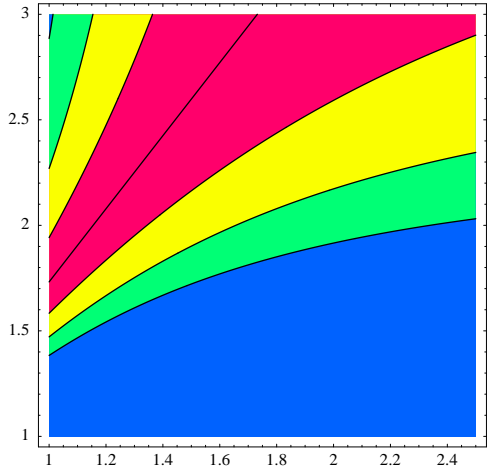,width=0.4\textwidth,angle=0}}
	\caption[sniff]{Contours of constant $\delta$
	  in the $\gtp$--$\gt$ plane (see text).  
	  On the $\gt = \sqrt{3}\gtp$
	  line there is no correction ($\delta=0$).  
	  Going outward from this line are regions with 
	  $\delta <2\%$ (red), $<4\%$ (yellow), $<6\%$ (green), 
	  and $>6\%$ (blue).}
\label{precision}
\end{figure}

%
To measure the sensitivity of $\sint$ to $\gt$ and $\gtp$, 
we fix $M$ to be the value computed in the previous section,
$M_o=3.75$ TeV.  We then define 
\begin{eqnarray}
\delta\equiv\frac{\delta\sint}{1/4}
   &=& \frac{3-\frac{g^2}{g'^2}}{1+\frac{g^2}{g'^2}} \nonumber\\
   &=& \frac{g(M)^2 \left( \frac{3}{\gtp^2} - \frac{1}{\gt^2} \right)}{4 - 
	g(M)^2 \left( \frac{3}{\gtp^2} - \frac{1}{\gt^2} \right)}
\end{eqnarray}
where $g(M)$ is obtained by running the measured value of $g$ from 
$\Mz$ to $M$.  The $\delta$ is the fractional deviation of 
$\sint$ from its $SU(3)$-symmetric value of $.25$ at $M$.  
In Figure \ref{precision}, we plot values of $\delta$ as a function
of $\gt$ and $\gtp$.  We limit the upper end of the parameter space
to avoid the strong-coupling regime and Landau pole (we discuss the
latter below).  We see from the figure that in a significant portion 
(about 25\% of the plotted region) of the parameter space 
with $\gt,\gtp>1$ the deviation of $\sint$ from its $SU(3)$ value
is less than $2\%$.  

Adding renormalization group running to eqs. (\ref{geq}) and (\ref{gpeq})
produces
\begin{equation}
{8\pi^2\over g(\Mz)^2} = {8\pi^2\over g_3^2} 
	- b \ln{\frac{M_o}{\Mz}} 
	- b \ln{\frac{M}{M_o}} 
	+ {8\pi^2\over \gt^2}\, ,
\label{geq2}
\end{equation}
and
\begin{equation}
{8\pi^2\over g'(\Mz)^2} = {8\pi^2\over g_3^2/3} 
	- b' \ln{\frac{M_o}{\Mz}} 
	- b' \ln{\frac{M}{M_o}} 
	+ {8\pi^2\over \gtp^2}\, .
\label{gpeq2}
\end{equation}
We take the last two terms as the source of theoretical uncertainty of 
the module, where we have used $M_o$ as our ``central value''.  Note
that for $M$ in the range 1--10 TeV, its contribution to the fractional
uncertainty is $\sim 3\%$.  Eliminating $g_3$, we can
find the $SU(3)$-breaking scale $M$ as a function of $(\gt,\gtp)$:
\begin{equation}
M = \Mz e^{\frac{8\pi^2}{b-b'/3} 
	[(\gt^{-2} - \gtp^{-2}/3) - (g^{-2} - g'^{-2}/3)]}
\label{Mfunc}
\end{equation}
where again the SM gauge couplings are evaluated at $\Mz$.

For obvious reasons, we do not extend the parameter space of the tilded
gauge couplings to infinity.  Even at semi-perturbative levels, there
is an additional degree of uncertainty introduced due to the existence 
of a Landau pole in the $\gtp$ coupling.  This puts an upper limit on an 
ultraviolet cutoff $\Lambda$ in the theory.  Unknown ultraviolet physics 
parameterized by operators such as 
\begin{equation}
\frac{|\Sigma|^2}{\Lambda^2} F^{\mu\nu}F_{\mu\nu}\, ,
\label{operator}
\end{equation}
where $F^{\mu\nu}$ is the weak or hypercharge gauge bosons, 
can change the predicted value of $\sint$ at $M$.
These effects limit the parameter space in the region of
large $\gtp$.  

The location of the Landau pole is defined as the scale
at which the coupling $\gtp$ blows up from the effects
of one-loop running:
\begin{equation}
\Lambda = M e^{-\frac{8\pi^2}{b'_{\Sigma} \gtp^2}}
\end{equation}
where the beta-function coefficient $b'_{\Sigma} =  -22/3$ includes
the $\Sigma$ field in the $(3,2_{-1/2})$ representation.
For $\gtp(M)=2.2$, the Landau pole is an order of magnitude
above $M$ and operators like (\ref{operator}) contribute less than
1 \% uncertainty.  For $\gtp(M)=3.0$ the Landau pole is only a factor
of 3.3 above $M$ and physics at the scale $\Lambda$ may affect the
prediction of $\sint$.

Embedding the $U(1)$ into an asymptotically free group at $\sim M$ avoids
the issue of the Landau pole.  We discuss two such possibilities in the
next section.

There are additional uncertainties coming from the normal 
threshold effects which will contribute to $\sint$ as well.  
In the minimal model, the unknown effects come from the 
Higgs and $\Sigma$ masses.  For weak-scale Higgs masses, 
the Higgs threshold correction to $\sint$ is estimated to 
be less than two tenths of a percent \cite{lp}.  The $\Sigma$ 
contribution can be estimated by allowing the mass of the 
remaining physical triplet (see the Appendix) to vary over 
a range of scales.  It too has a relatively small effect
for masses between $1$ and $10$ TeV.  

Now let us compare to the SSM.
Although the SSM has 125 parameters
\cite{Dimopoulos:1995ju}, its gauge sector (at one loop) is not contaminated
by the vast flavor sector. This leads to the prediction of the
weak mixing angle as follows. The experimental values of the three
gauge couplings at $\Mz$ are determined by three theoretical
parameters: the unification scale $M_{GUT}$, the value of the common
gauge coupling $g_{GUT}$ at that scale, and the scale of superpartner
masses $M_s$. Since the number of experimental and theoretical
parameters are the same, we expect no real prediction. However,
for small $M_s$, not too far from $\Mz$, the prediction of
$\sint$ is sensitive to the value of $M_s$ only at the
percent level.  So, for a range of relatively small $M_s$, we obtain
the successful $\sint$ prediction. It is, perhaps, more
correct to view this as a prediction of the presence of light
sparticles near  $\Mz$, rather than a prediction of
$\sint$. A reason why this is not viewed as ``fitting
$\sint$ by adding $\sim 100$ new particles'' is that
the presence of sparticles near $\Mz$ is independently motivated
by the hierarchy problem.

In our present proposal the accounting is similar. 
For $\tilde{g}$, $\tilde{g{\prime}}$ in the broad range shown in
the figure, the value of  $\sint$ is independent of
these at the few \% level and is determined by the
$SU(3)$-symmetric dynamics encoded in $g_3$.  
Thus the experimental values of the two electroweak gauge couplings are
determined in terms of the two parameters $g_3$ and $M$. 
For small $M$ not too far from $\Mz$, the prediction of $\sint$
depends logrithmically on, and is therefore insensitive to $M$.
Just as in the SSM, we can view the smallness of the theoretical
uncertainty, perhaps more correctly, as a prediction 
of concrete new physics at the scale M$\sim few$ TeV.

A fundamental distinction is that the SSM is an explicit {\it
model}, whereas here we are proposing a {\it mechanism} (or {\it
module}) which may be embedded in different models.
The simplest models implementing our mechanism have a fine tuning
of about one part in $10^3$ or $10^4$ to ensure that the Higgs
remains lighter than the fundamental scale. In the SSM the fine
tuning inferred from the absence of light sparticles is 
about 1\%.  Combining supersymmetry with our mechanism would reduce 
the fine tuning to the same 1\% level.  We discuss another possibility
in a later section.

\noindent{\bf Charge Quantization:}
One prediction common to all semi-simple grand unified theories is
quantized hypercharge.  The model described above contains a $U(1)$
factor and therefore does not require charges to be rational.  The prediction
of $\sint$ at $M$ is in fact dependent on a {\it continuous}
parameter, the $U(1)$ charge of $\Sigma$.  For a charge $-x$ in the 
same canonical normalization of SM hypercharges, the unbroken $U(1)$ gauge
boson below the scale $M$ is 
\begin{equation}
B_{\mu} = 
\frac{\gtp x A^8_{\mu} + (g_3/2\sqrt{3}) {\tilde B}_{\mu}}{\sqrt{g_3^2/12
	+ \gtp^2 x^2}}\, ,
\end{equation}
where $A^8_{\mu}$ and ${\tilde B}_{\mu}$ are the $SU(3)$ ``hypercharge''
and $U(1)$ gauge bosons respectively.  A SM field $\psi$ with $U(1)$ charge
${\tilde Y}$ will couple to $B_{\mu}$ as
\begin{eqnarray}
i {\tilde Y} \gtp {\tilde B}_{\mu} \psi 
&=& \frac{i {\tilde Y} \gtp g_3/2\sqrt{3}}{\sqrt{g_3^2/12 + \gtp^2 x^2}} B_{\mu}
+ \cdots \nonumber\\
&\rightarrow& i {\tilde Y} \frac{g_3}{2\sqrt{3} x} B_{\mu} + \cdots \nonumber\\
&\equiv& i Y g' B_{\mu} + \cdots
\end{eqnarray}
where in the second line the limit $\gtp\rightarrow\infty$ is taken.
Having chosen charges such that $Y={\tilde Y}$, we see that 
$g'=g_3/(2\sqrt{3} x)$ and 
\begin{equation}
\sint = \frac{1}{1+12x^2} \:\:\: {\rm at}\:\: M\, .
\end{equation}
For continuous $x$ there exists a continuous set of predictions for
$\sint$.  However, if the scenario is embedded in a more fundamental theory 
where charge quantization is generic ({\it i.e.}, string theory),
the prediction of $\sint$ takes on more significance.
The model may also be embedded in a field theory which predicts 
rational charges, such as a semi-simple Yang-Mills theory.  
Since the larger group manifests itself at low energies,
it must not mediate proton decay.  Two well known examples 
of such a group are Pati-Salam 
($SU(4)_c\times SU(2)_L \times SU(2)_R$) \cite{Pati:1974yy}
and trinification ($SU(3)_c\times SU(3)_L\times SU(3)_R/\cZ_3$) 
\cite{tri}.

In Pati-Salam, there are two independent gauge couplings when the 
two $SU(2)$ coupling are set equal by imposing parity.  The field 
content of the SM fits beautifully:  the matter multiplets are the 
$(4,2,1)$ containing the quark and lepton doublets, the $({\bar 4},1,2)$ 
containing the quark and lepton singlets (including a right-handed neutrino), 
and a scalar $(1,2,2)$ containing two Higgs doublets.  Under the 
Pati-Salam group, $\Sigma$ must transform as a multiplet which contains 
a doublet of charge 1/2, {\it e.g.}, $(1,2,2)$ or $(4,2,1)$.  

The $SU(2)\times U(1)$ can also be embedded into $SU(3)^3$ 
(without the modded out $\cZ_3$ factor).  The gauge couplings of the 
$SU(3)_{L,R}$ portion can be made relatively strong 
(and equal by again imposing 
a parity symmetry) and the standard multiplets contain the 
SM fermions as the only chiral matter.  The $\Sigma$ can be a 
$(3,1,3,{\bar 3})$ under ``quadrification'':  $SU(3)\,$-new, color,
left and right respectively.  

In both Pati-Salam and quadrification additional breaking of gauge symmetries
must occur.  This can be done in the usual way for these groups \cite{Pati:1974yy,tri}
or may be incorporated with the breaking of the weak $SU(3)$.
Some additional structure is required to avoid unwanted Yukawa relations
and (too-)large neutrino masses.  For example, in quadrification, 
fermion masses are unrelated if there are two Higgs multiplets - one for 
quarks and one for leptons.  A complete description of models of this type
will appear in future work \cite{big}.

Embedding the $U(1)$ group of the minimal module into semi-simple ones
tend to widen the parameter space to which $\sint$ is insensitive.  
This is because after breaking the semi-simple group down to the 
strong $SU(2)\times U(1)$, the range of values for $\gtp/\gt$ 
is smaller than the normal range (zero to infinity) due to the 
correlation of $\gtp$ to other gauge couplings.
In Pati-Salam, for instance, the effective $U(1)$ coupling $\gtp$ 
is a function of $g_4$ and $g_R$:
\begin{equation}
\gtp = \frac{g_R g_4 \sqrt{3}/(2\sqrt{2})}{\sqrt{g_R^2/4 + 3 g_4^2/8}}
\end{equation}
where $g_4$ at the breaking scale $M_{PS}$ is equal to the QCD coupling
$g_c$ at $M_{PS}$, which can be determined by running its measured value
up from $\Mz$.  
Note that for $g_R\rightarrow\infty$, $\gtp$ does not blow up but
hits the asymptotic value of $\sqrt{3/2} g_4$.  
In quadrification, $\gtp$ is a 
function of the left and right $SU(3)$ couplings:
\begin{equation}
\gtp = \frac{g_L g_R/2}{\sqrt{g_R^2/12 + g_L^2/3}}
\end{equation}
Here in the limit $g_R\rightarrow\infty$, $\gtp\rightarrow \sqrt{3} g_L$.
In neither case does the ratio $\gtp/\gt$ blow up anywhere in the 
parameter space.

\begin{figure}
  \centerline{
    \psfig{file=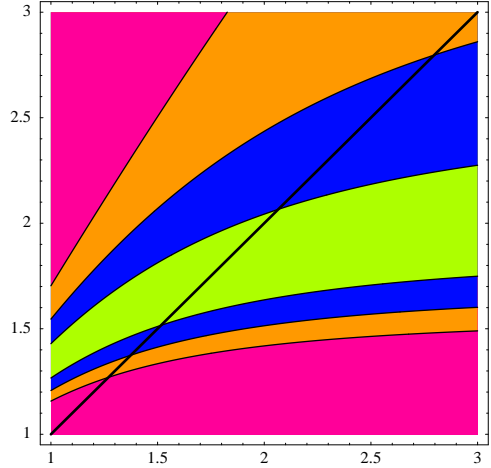,width=0.4\textwidth,angle=0}}
	\caption[sniff]{Contours of constant $\delta$
	  in the $g_R$--$g_L$ plane for Pati-Salam.  
	  The thick black line corresponds to the parity-symmetric case.  
	  Going outward from the central region are regions with 
	  $\delta <2\%$ (green), $<4\%$ (blue), $<6\%$ (orange), 
	  and $>6\%$ (magenta).}
\label{PS}
\end{figure}

\begin{figure}
  \centerline{
    \psfig{file=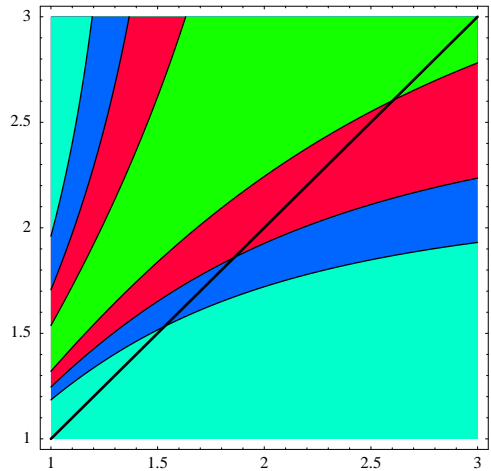,width=0.4\textwidth,angle=0}}
	\caption[sniff]{As in Figure \ref{PS}, but for quadrification.
	  Again the parity-symmetric case is shown.  The colors
	  outward from the central region are 
	  $\delta <2\%$ (green), $<4\%$ (red), $<6\%$ (blue),
          and $>6\%$ (aqua).}
\label{Quad}
\end{figure}

Figures \ref{PS} and \ref{Quad} are the analogs of Figure \ref{precision}
for Pati-Salam and quadrification respectively.  They are plotted in the
plane of the $SU(2)_L \times SU(2)_R$ and $SU(3)_L \times SU(3)_R$ 
gauge couplings where we have dropped the parity restriction.  
We are not working from explicit models and for simplicity we 
assume that the larger group breaks to the minimum module at the
same scale $M=3.75$ TeV as the weak $SU(3)$.

We see from the figures that the effects on $\sint$ are small 
in a larger segment of the parameter space. 
Restoring parity as a symmetry 
restricts the parameter space to the diagonal line on each plot.  
On the parity-symmetric line in Figure \ref{PS} $\sint$ is particularly
insensitive -- much of the line lies in the $< 4\%$ correction region.

The inclusion of the module into a semi-simple group produces 
three benefits: 1. charge quantization, 2. removal of the Landau pole,
and 3. enhancement of the insensitive region of parameter space.
The examples we have presented are the well known groups 
into which SM matter fits perfectly, {\it i.e.}, is the only chiral
matter in the complete multiplets.    Thus we are able to incorporate this 
success of standard GUTs.  Constructing the complete models, taking into
account all phenomenological consequences, would be a worthwhile endeavor.

\noindent{\bf The Hierarchy Problem:}
As mentioned before, the minimal module has a fine tuning of at least
one part in $10^3$ associated with the Higgs mass.  Of course, the 
module can easily be supersymmetrized to reduce the fine tuning.  
The purpose of this section is to present an example in which the
hierarchy problem is ameliorated without the use of supersymmetry while
maintaining the successful prediction of $\sint$.  For our example, 
we choose a model in which the Higgs is a pseudo-Goldstone boson
\cite{Kaplan:1983fs,Arkani-Hamed:2001nc}.  We present a simple 
version of these models which leaves a fine-tuning 
of about $\sim 1\%$.  Versions with more structure along the lines 
of \cite{Arkani-Hamed:2001nc} can eliminate the fine-tuning 
completely.

The model has two sets of Nambu-Goldstone bosons which can be written
as a non-linear sigma model in the standard form \cite{weak}:
\begin{eqnarray}
U&=&e^{i \pi_u/2f}\nonumber\\
V&=&e^{i \pi_v/2f},
\end{eqnarray}
where for simplicity we have set the decay constants $f$ equal.
Each multiplet transforms under an independent $SU(4)_L\times SU(4)_R$ chiral
symmetry:
\begin{equation}
U\rightarrow L U R^{\dagger}
\hspace{0.25in}
V\rightarrow {\cR} V {\cL}^{\dagger}.
\end{equation}
The ``pions'' transform linearly (in the adjoint representation)
under the symmetric combination of their
respective chiral symmetries and non-linearly under the anti-symmetric
combinations.

\begin{figure}
  \centerline{
    \psfig{file=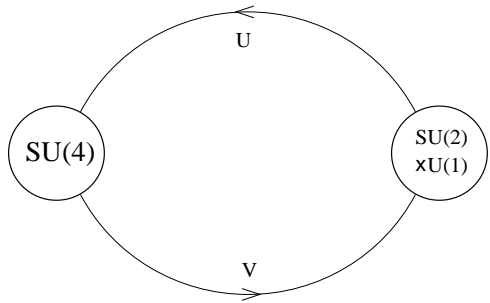,width=0.4\textwidth,angle=0}}
	\caption[sniff]{A diagrammatic description of the pseudo-Goldstone
	  boson Higgs model.  The two lines represent Goldstone bosons in
	  the $(4,{\bar 4})$ and $({\bar 4},4)$ of two sets of
	  $SU(4)_L\times SU(4)_R$ chiral symmetries.
	  The circles represent the gauging of the diagonal linear 
	  combination of the $SU(4)_L$ symmetries and an $SU(2)\times U(1)$
	  subgroup of the $SU(4)_R$ symmetries.}
\label{moose}
\end{figure}

The chiral symmetries are broken by the following perturbations:
\begin{enumerate}
\item The~diagonal combination of the two $SU(4)_L$ global symmetries is
weakly gauged.
\item An $SU(2)\times U(1)$ subgroup of the diagonal combination of $SU(4)_R$
global symmetries is gauged somewhat more strongly than the $SU(4)_L$.  
\end{enumerate}
The model is represented by the diagram in Figure \ref{moose}.

Both $U$ and $V$ break the gauged $SU(4)$ and $SU(2)\times U(1)$ to a diagonal
$SU(2)\times U(1)$ at the scale $f$.  The $U(1)$ is chosen\footnote{This may
be a natural consequence of embedding the $SU(2)\times U(1)$ in a 
semi-simple group, {\it e.g.}, Pati-Salam, broken at the cutoff and 
imposing some discrete symmetries.} such that the $SU(2)\times U(1)$ 
live in an $SU(3)$ subgroup of the diagonal $SU(4)_R$.  Under the 
diagonal (fictitious) $SU(3)$ the pions transform as an
$8+3+{\bar 3}+1$.  Choosing Higgs doublets with hypercharge $\pm 1/2$
to live in the $3 + {\bar 3}$ gives the prediction 
$\sin^2{\theta_W} \sim .25$ at the scale $f$.  

Now it is clear why we had to choose $SU(4)$ instead of $SU(3)$.
The pseudo-Goldstone bosons come in adjoint representation of 
the vector part of the chiral symmetry.  We require a group which contains 
an $SU(3)$ subgroup and whose adjoint representation contains a triplet.  
The smallest possibility is $G_2$ whose adjoint is a 14 and contains 
$8+3+{\bar 3}$.  Because it is rank 2 there are discrete choices for
$SU(2)\times U(1)$.  However, for the simplicity of this description,
we have chosen the example more indicative of QCD.

The $SU(4)$-adjoint pions break down under 
$SU(2)\times U(1)$ to an uncharged triplet, two uncharged singlets, a complex 
charge-one singlet, a complex charge-3/2 doublet and a complex charge-1/2 
doublet.  The last of these can play the role of the SM Higgs.  

The fate of the two sets of pions is as follows:  one linear combination
is eaten by the breaking of the gauge symmetry to the diagonal group while
the other can be parameterized in unitary gauge as
\begin{equation}
U*V\equiv {\cal U} = e^{i \pi/2 f}\, .
\end{equation}
Note the scale $f$ is where $\sint$ exhibits the approximate $SU(3)$ relation.
Therefore we take $f\sim few$ TeV.  The gauge couplings explicitly break 
the global symmetries of the theory and distinguish between different
components of the pions.  Therefore, we expect to generate operators at 
one loop which treat the remaining Goldstones differently.
Some examples of operators are:
\begin{eqnarray}
&{\rm tr}\,[\cU T^a \cU^{\dagger} T^a]&
\hspace{0.25in}
|{\rm tr}\,[\cU T^a]|^2
\nonumber\\
&{\rm tr}\,[\cU T^8 \cU^{\dagger} T^8]&
\hspace{0.25in}
|{\rm tr}\,[\cU T^8]|^2
\end{eqnarray}
where the $T\,$s are $SU(4)$ generators and $a = 1,2,3$.
These operators will give both quadratic and quartic couplings
to the pions, with the different $SU(2)\times U(1)$ representations
getting different masses.  In addition, there will be couplings
to fermions, most importantly the top quark, which distinguish
between the pseudo-Goldstone bosons.  It is hopeful that in some 
region of parameter space, the charge-$1/2$ doublet gets the 
(dominant) VEV and properly breaks
electroweak symmetry.  The natural scale for the VEV is $\sim f$
and so fine-tuning of about $1\%$ is required to get the correct
electroweak symmetry breaking scale relative to the higher
$SU(3)$-breaking scale.

Additional structure like a two-by-two lattice of the type introduced 
in \cite{Arkani-Hamed:2001nc} can remove the fine tuning entirely 
\cite{big}.  In such a model, the size of the VEV is naturally a loop 
smaller than the decay constant, or in other words, $v \sim f/4\pi$.  
Remarkably, this is just what is needed to run from the $SU(3)$ value of 
$\sint$ to the measured value.  We postpone discussion of models of this 
type to a longer article \cite{big}.

%

\noindent{\bf Discussion:} In this framework, the standard
model is valid up to energies of order a few TeV.
At that scale new particles and phenomena begin to emerge; some
are model independent and an integral part of the mechanism we are
proposing.  These include the extra gauge bosons associated with
the full $SU(3)\times SU(2) \times U(1)$ as well as the $\Sigma$
multiplet that bridges the $SU(2) \times U(1)$ and $SU(3)$.  Also,
at a few TeV, the
$SU(2)\times U(1)$ forces become strong. 
What happens beyond that is model dependent. Some new physics
must set in to protect the theory from the $U(1)$ Landau pole and
to account for charge quantization. Our favorite possibility is
that full string theory emerges at a few TeV. Other, more explicit
possibilities -- such as Pati-Salam, quadrification or composite
Higgs models-- have already been discussed in previous sections.
In each case there is rich new physics to look forward to.

There are at least three other
approaches to computing  $\sint$ in theories with
string scale near a TeV. The earliest 
\cite{Dienes:1998vh} relies on
accelerating the normal evolution of gauge couplings \cite{dg} by
having TeV$^{-1}$ size extra dimensions. A potential obstacle to
this approach is that the values of higher dimensional gauge
couplings are very sensitive to unknown short-distance physics.
Another approach, valid for two large dimensions, exploits the
mapping between ordinary RG-evolution and the profile of 2-D bulk
fields 
It would be interesting to construct a realistic
model based on this. Another proposal 
\cite{Arkani-Hamed:2001vr} introduces $\sim 13$ 
copies of the gauge and Higgs sector of the SSM. This
reproduces the prediction of the SSM, though with significant
theoretical uncertainty $\sim 20\%$ -- due to the threshold
effects from each of the sectors.

These three proposals rely essentially on the same numerical
reasons as the SSM to reproduce the value for $\sint$.
In this paper we presented a different way to derive the value of
$\sint$, unrelated to the SSM. It involves a low-energy
$SU(3)$ symmetry whose presence can be directly tested at the LHC.

{Acknowledgments:} We are happy to thank Nima Arkani-Hamed,
Lawrence Hall, David B. Kaplan, Steven Shenker and Yonathan Zunger 
for valuable discussions. This work is supported by 
NSF grant PHY-9870115 and DOE grant DE-AC03-76SF00515.

\noindent{\bf Appendix -- Diagonal Breaking and the Potential:}
The most general renormalizable potential for $\Sigma$,
\begin{equation}
{\cal{V}} = \lambda \left({\rm tr}\Sigma^{\dagger}\Sigma - v^2\right)^2
- \lambda'\,{\rm det}\,\Sigma^{\dagger}\Sigma + {\rm const.}
\label{potential}
\end{equation}
can produce the correct symmetry breaking
$SU(3)\times SU(2)\times U(1)\rightarrow SU(2)\times U(1)$.
To see this -- and the resulting spectrum --
we parameterize $\Sigma$ as follows:
\begin{eqnarray}
\Sigma = e^{i \pi^a \lambda^a/2 M}\times\\ \nonumber
\pmatrix{(\eta_0 + M) + \eta_3&\eta_1 - i\eta_2 \cr
        \eta_1 + i\eta_2&(\eta_0 + M) - \eta_3 \cr 0&0}
        e^{i \pi^{a'} \sigma^{a'}/2 M}\, ,
\end{eqnarray}
where $M$ is the scale of symmetry breaking and $\lambda$, $\sigma$
are Gell-Mann and Pauli matrices respectively, $a=1-8$ and $a' = 1,2,3,8$.
We identify $\sigma^8$ with the identity matrix of size two.  Of the twelve
degrees of freedom in the complex $\Sigma$, eight are Nambu-Goldstone
bosons (the $\pi\,$s) eaten by the broken gauge generators.  The potential
for the remaining Higgs fields is
\begin{eqnarray}
{\cal{V}} = & \lambda \left((\eta_0 + M)^2 + {\eta}^2 - v^2\right)^2 
	\nonumber\\
        -&\lambda'\,\left((\eta_0 + M)^2 - {\eta}^2\right)^2,
\end{eqnarray}
with $\eta = \{\eta_1,\eta_2,\eta_3\}$.  If $\lambda>\lambda'>0$, there is 
a discrete minimum for which
$M = \sqrt{\lambda/(\lambda - \lambda')} v$ and
a VEV of the form (\ref{vev}) is reproduced.  The $SU(2)$ triplet $\eta$ 
gets a positive squared mass at tree level of size 
$m^2 = 4\lambda\lambda' v^2/(\lambda - \lambda')$.  The singlet $\eta_0$
stays massless at tree level, but has a quartic coupling which induces
a mass at one loop.

\end{document}